\documentclass[12pt]{article}
\usepackage[utf8]{inputenc}
\usepackage[margin=1in]{geometry}
\usepackage[
backend=biber,
sorting = none,
]{biblatex}
\usepackage{hyperref}
\hypersetup{
    colorlinks=true,
    linkcolor=blue,
    filecolor=magenta,      
    urlcolor=cyan,
    citecolor= red
}

\usepackage{parskip}
\usepackage{graphicx} 
\usepackage{float} 
\usepackage{amsmath}
\usepackage{xcolor}
\usepackage{blkarray} 

\renewcommand*{\thesection}{\arabic{section}}

\begin{document}

\begin{center}
    \large \textbf{Emergence of simple and complex contagion dynamics from weighted belief networks}
    \\\small Rachith Aiyappa$^{a,1}$, 
    Alessandro Flammini$^{a}$, 
    Yong-Yeol Ahn$^{a,}$\footnote{$^1$ To whom correspondence should be addressed. E-mail: racball@iu.edu and yyahn@iu.edu} \\
    $^a$Center for Complex Networks and Systems, \\
    Luddy School of Informatics, Computing, and Engineering, \\
    Indiana University, Bloomington, Indiana, USA, 47408
\end{center}
\bigskip

\textbf{Link to the paper published in Science Advances:} \url{https://www.science.org/doi/10.1126/sciadv.adh4439}
\\

\textbf{BibTeX}
\color{red}
\small
\begin{verbatim}
@article{
aiyappa2024emergence,
author = {Rachith Aiyappa and Alessandro Flammini and Yong-Yeol Ahn},
title = {Emergence of simple and complex contagion dynamics from weighted belief networks},
journal = {Science Advances},
volume = {10},
number = {15},
pages = {eadh4439},
year = {2024},
doi = {10.1126/sciadv.adh4439},
URL = {https://www.science.org/doi/abs/10.1126/sciadv.adh4439},
eprint = {https://www.science.org/doi/pdf/10.1126/sciadv.adh4439}
}
\end{verbatim}

\color{black}
\noindent 
\section*{Abstract}\label{sec:abstract}

Social contagion is a ubiquitous and fundamental process that drives individual and social changes. 
Although social contagion arises as a result of cognitive processes and biases, the integration of cognitive mechanisms with the theory of social contagion remains an open challenge. 
In particular, studies on social phenomena usually assume contagion dynamics to be either simple or complex, rather than allowing it to emerge from cognitive mechanisms, despite empirical evidence indicating that a social system can exhibit a spectrum of contagion dynamics---from simple to complex---simultaneously. 
Here, we propose a model of interacting beliefs, from which both simple and complex contagion dynamics can organically arise. 
Our model also elucidates how a fundamental mechanism of complex contagion—resistance---can come about from cognitive mechanisms.
\section{Introduction}\label{sec:introduction}
From the spreading of infectious diseases to the diffusion of ideas, innovations, beliefs, and behaviors, social contagion is a fundamental social dynamic that drives large-scale changes in societies~\cite{bond201261,coleman1957diffusion,anderson1992infectious,easley2010networks,rogers2014diffusion}. 
It is at the heart of numerous social challenges that our society is facing~\cite{lorenz2022systematic}, including the prevalence of false information~\cite{lazer2018science}, political polarization~\cite{conover2011political,macy2019opinion}, climate change denial~\cite{gronow2021policy}, and vaccine hesitancy~\cite{jolley2014effects,broniatowski2018weaponized}.
These challenges are tied to the deluge of information being exchanged in the modern world, particularly via social media, that offers frictionless conduits for any kind of information to spread globally.
This accelerated spread and efficient discovery process may be exacerbating our cognitive biases and flawed cognitive decision-making system~\cite{kunda1990case,hornsey2020facts}. 

Various cognitive biases come into play when people encounter new information and decide to share it or not~\cite{kahneman2011thinking,ciampaglia2018biases}. 
Although they may often be useful heuristics and shortcuts, cognitive biases like confirmation bias~\cite{nickerson1998confirmation}, can also lead us to discard or misinterpret facts and evidence, especially when it is complex or contradictory to our existing beliefs~\cite{festinger2017prophecy}. 
This tendency can thus aggravate the polarization and spread of false information in society. 

\begin{figure}
    \centering
    \includegraphics[width=1\textwidth, trim={0 0in 0 0in}]{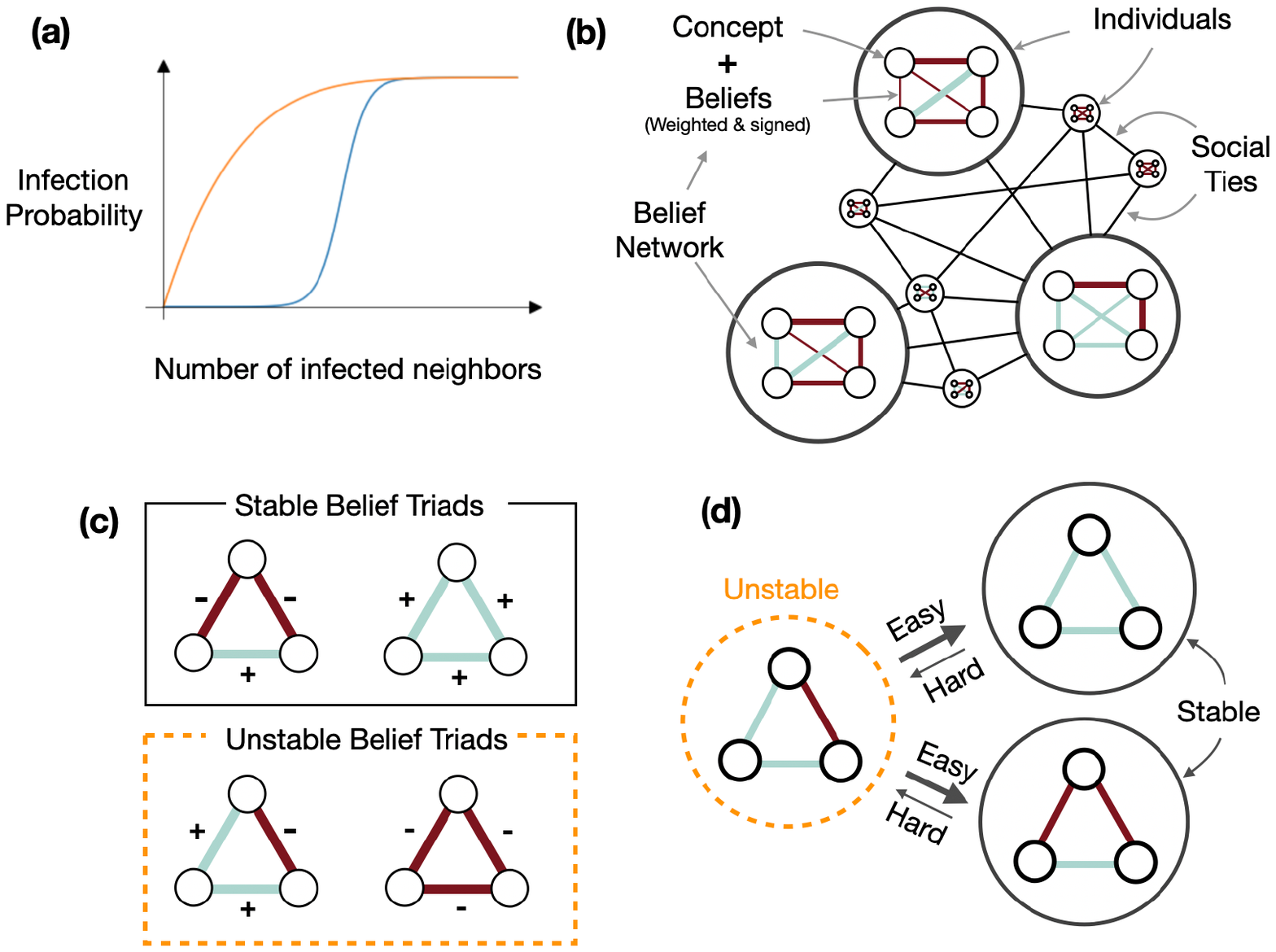}
    \caption{(a) An illustration of the simple and complex contagion mechanisms, as reflected in the adoption probability’s dependence on the number of adopted neighbors. 
    Simple contagion (orange curve) is characterized by a concave shape and exhibits ``diminishing returns'' due to the independence of the exposure’s impact. 
    On the other hand, the complex contagion curve (blue curve) shows a sharp increase due to the reinforcement. 
    (b) An individual’s belief system is described as a network of interacting beliefs, where nodes represent concepts and weighted edges between them represent beliefs. 
    Beliefs describe the (personal) degree of coherence that agents attribute between pairs of concepts. 
    The color of the edge indicates the belief polarity and the thickness indicates its strength. 
    (c) The stability of each triad in an individual’s belief network is modeled using the social balance theory.
    (d) Each belief network has a bias toward internal coherence (stable triads).}\label{fig:setup_dynamics}
\end{figure}

On the modeling side, the dynamics of social contagion are typically studied with a social network where a ``susceptible'' node may be infected by ``infected'' neighbors, and where the likelihood of such an event is usually assumed to be a function of the number of exposures. 
The function that determines contagion dynamics falls into two broad classes—simple and complex contagion~\cite{centola2007complex,centola2010spread,romero2011differences,weng2013virality,dodds2004universal,dodds2005generalized,porter2016dynamical}.
Although stringent and universally agreed-upon definitions of simple and complex contagion have not yet been established in the literature, the difference can be explained with representative examples. 
An illustrative example of simple contagion is the infection process in the SIR model~\cite{anderson1992infectious}, or the independent cascade model~\cite{goldenberg2001talk}, where each infected neighbor contributes independently to the probability of infection of a target node. This leads to a concave shape of the infection probability curve—probability of infection versus the number/fraction of infected nodes (see Fig.~\ref{fig:setup_dynamics}a)---characterized therefore by diminishing returns with respect to an increasing number of infected neighbors. 
By contrast, in complex contagion, the contribution of the infected neighbors is not independent anymore and the social reinforcement becomes critical. 
Namely, when there is little (not enough) exposure, the probability of adoption remains low; however, as we cross the threshold of adoption, the probability increases markedly, producing an S-shaped adoption curve (see Fig.~\ref{fig:setup_dynamics}a)~\cite{centola2007complex,goldenberg2001talk,granovetter1978threshold,watts2002simple}, that is often approximated with a step function~\cite{granovetter1978threshold,watts2002simple}. 
Rather than depending on the specific knowledge of the infection mechanisms [which may be difficult to acquire in many application settings~\cite{hebert2020macroscopic}], for the sake of the present paper, we will use the shape of the infection curve as a discriminating factor between simple and complex contagion.

These two classes of dynamics reflect distinct characteristics of contagion, such as the adoption of costly behaviors (e.g., a healthy diet) versus cheap ones (e.g., rumor), and have profound implications on the patterns of spreading and how such spreading depends on the social network structure~\cite{centola2007complex,centola2010spread,redlawsk2010affective, nematzadeh2014optimal}. 
For instance, it has been shown that, although inter-community bridges unilaterally facilitate simple contagion~\cite{weng2013virality,shirley2005impacts}, complex contagion can both be facilitated and inhibited by inter-community bridges~\cite{morris2000}, exhibiting a more nuanced behavior~\cite{nematzadeh2014optimal}. 

Many models of contagion have been proposed to account for the phenomena like polarization and consensus formation~\cite{castellano2009statistical}.
However, social contagion models rarely consider the psychological/cognitive basis of social contagion. Instead of starting from cognitive mechanisms to derive contagion dynamics, studies assume dynamics and do not account for complex belief interactions within individuals, although these interactions could be integral to social contagion and responsible for counterintuitive phenomena
~\cite{dalege2022using,miton2015cognitive,kahan2011cultural}.
A wide gap still exists between models of cognitive processes and models of social contagion~\cite{sobkowicz2009modelling}, although the effect of the human predisposition to have a coherent set of beliefs~\cite{festinger1962cognitive} on social contagion has been of recent interest~\cite{goldberg2018beyond,houghton2020interdependent,galesic2021integrating,rabb2022cognitive,friedkin2016network}. 
In particular, it would be highly desirable to have a theoretical model that bridges the gap and derives the emergence of both complex and simple contagion and, possibly, a range of observed intermediate behaviors~\cite{dodds2004universal} from the same theoretical framework.

Here, we propose a belief dynamics model that couples the cognitive tendency toward internal coherence with social contagion theory. Our model builds on recent work that models the internal belief system as an undirected, unweighted network of semantic concepts~\cite{rodriguez2016collective}.
We introduce the strength of beliefs and natural dynamics that account for the tendency to have a coherent set of beliefs and to align beliefs with those of the people one interacts with. We show that, depending on the context in which beliefs exist, both simple and complex contagion dynamics emerge organically from the dynamics of the belief network.

\section{Interacting weighted beliefs model}\label{sec:model}

Consider a social network of $N$ nodes and $M$ edges. 
Each node represents an individual ($i$) and each edge represents a social relationship through which (the strength of) beliefs are exchanged. 
Each individual's belief system is described via \emph{another} network---the \emph{belief network}, $B^i$. In the belief network nodes represent entities (e.g., Roger Federer and London), concepts (e.g., vaccinations and abortion rights), or notions (e.g., good and dangerous). 
For the sake of simplicity, in this paper, we abstract out these differences and treats all nodes of the belief network the same, although in principle different types of nodes can be explicitly modeled.
The edge between a pair of nodes represents, from an individual perspective, how coherent (or mutually exclusive) are the notions of the corresponding nodes. 
We refer to edges as ``\emph{beliefs}'', denoted by $b$. 
A belief is therefore represented by an undirected \emph{weighted} edge, where its sign reflects the belief's polarity and its weight (ranging in $[-1,1]$) represents the strength of the belief (see Fig.~\ref{fig:setup_dynamics}b). 
In our model beliefs change as a consequence of social interactions and because of the predisposition to hold a coherent set of beliefs~\cite{festinger1962cognitive}. 

Let us illustrate the notion of the belief network and dissonance/coherence with a toy example about a fictional agent, Bob.
Bob is an avid follower of tennis, who detests \textit{match fixing}. \textit{Roger Federer} has been his favorite tennis player for a long time. 
Thus, in Bob's belief network, there is a positive link between \textit{Federer} and \textit{good} (reflecting his affinity to \textit{Federer}) and a negative link between \textit{match fixing} and \textit{good}. There is also a negative link between \textit{Federer} and \textit{match fixing} (indicating Bob's belief that \textit{Federer} is not involved in any \textit{match fixing}). 
Now consider a hypothetical situation in which \textit{Federer} has been accused of \textit{match fixing}. 
This is reflected in a positive link between \textit{Federer} and \textit{match fixing} in the belief system of some of Bob's neighbors, whose influence can change Bob's belief system. 
Note that the information from the neighbors is not congruent with Bob's current belief system. Bob's beliefs that Federer is a good person and that match fixing is bad cannot easily reconciled with the fact that Federer may have committed the misdeed. 
To maintain his coherence, therefore, he may resist adopting this belief or may adopt it but change one of his existing beliefs, e.g., that Federer is a good person. 
This example also highlights the benefit of our choice of modeling beliefs as edges in contrast to alternatives which view them as nodes~\cite{dalege2022using,galesic2021integrating,rabb2022cognitive}. 
With this choice, we allow for the sharing of the same nodes (concepts) between multiple beliefs from which coherence between beliefs organically emerges via the balance theory (see Fig~\ref{fig:setup_dynamics}c and Eq~\ref{eq:internal_energy}).

We model the predisposition towards internal coherence by defining the internal dissonance of individual \(i\)'s belief network similar to the model described in Rodriguez et al.~\cite{rodriguez2016collective}, namely
\begin{equation}\label{eq:internal_energy}
    D^{i} = -\frac{1}{|\mathcal{T}|}\sum_{\substack{x,y,z \in \mathcal{T}}}b_x^{i}b_y^{i}b_z^{i}
\end{equation}
where \(b_x^{i}\) is a belief \(x\) in the belief network of \(i\), and the sum is evaluated over all triads in the belief network, denoted by set \(\mathcal{T}\) and normalized by the total number of triads (\(|\mathcal{T}|\)).
The lower the internal dissonance the more stable the belief system is (see Fig.~\ref{fig:setup_dynamics}c,~d)~\cite{festinger1962cognitive,stieglitz2013emotions}. 

Social contagion arises as an interaction between two individuals where one communicates a belief to another. 
Although richer characteristics of social communication can be accommodated, here, for the sake of simplicity, the interacting couples are chosen uniformly at random. 

At each time step \(t\), a random individual (sender, node \(j\)) communicates a randomly chosen belief \(b_{x}^j\) from their internal belief system. 
A random neighbor (receiver, node \(i\)) of the sender, in response to this communicated belief,  updates their internal belief (\(b_{x}^{i}\)) as  
\begin{equation}\label{eq:update_rule}
    \begin{gathered}
        b_{x}^i(t+1) = b_{x}^i(t) + f(b_x^j(t),B^i(t)),
    \end{gathered}
\end{equation}
where \(f(b_x^j,B^i)\) is a function of the sender's belief (\(b^{j}_x\)) and the receiver's current belief system $B^i$ at time t. 
The tendency of the receiver for internal coherence is captured by the derivative of \(D^{i}\) with respect to their current belief (\(b^{i}_x\)). That is,
\begin{equation}\label{eq:competition}
    f(b_x^j(t),B^i(t))  \sim  \alpha b_{x}^{j}(t) - \beta\frac{\partial D^{i}(t)}{\partial b_x^{i}(t)},
\end{equation}
where \(\alpha\) and \(\beta\) are parameters of the model that govern the strength of social influence and of the individual's predisposition towards internal coherence, respectively.
The derivative of internal dissonance with respect to a focal belief is given by $\frac{\partial D^{i}}{\partial b_x^{i}} = -\frac{1}{|\mathcal{T}|}\sum_{\substack{x',y,z \in \mathcal{T}}}b_y^{i}b_z^{i}I_{x'=x}$ where the sum is over all triads in the belief system and $I$ is an indicator function which takes the value of $1$ only when $x'=x$ and $0$ otherwise.
The more stable triads a positive focal belief $x (+)$ is a part of, the lesser the value of $\frac{\partial D^{i}}{\partial b_x^{i}}$, indicating that for greater internal coherence (or smaller internal dissonance), the strength of the $x$ should increase---stable triads become more stable. 
In other words, the sign of the derivative of the internal dissonance with respect to a focal belief captures the direction in which the focal belief of the individual should change for its internal dissonance to decrease. 
A negative (positive) sign indicates that for the internal dissonance to decrease, the strength of the focal belief should increase (decrease).

Eq.~\ref{eq:update_rule} does not constrain the strength of beliefs and thus the belief strength can diverge. To prevent it, we constrain the strength of each belief in the range $[-1, 1]$.
At each step of the simulation, after updating the belief strength, we set $b^i = \text{min} (1, b^i)$ if $b^i \geq 0$ and $b^i = \text{max} (-1, b^i)$ if $b^i \leq 0$.
A softer, but more complex, mapping function (for example, a sigmoid function) can also be used.~Eq.~\ref{eq:update_rule}, in unison with Eq.~\ref{eq:competition}, can lead to a belief $x$ of an individual $i(b^i_x)$ going to $0$.
However, this is not a case that requires special treatment and is in general only a transitory state. 
Unless the contribution of all triads of which $x$ is a part exactly balanced to $0$, the derivative of the dissonance (the tendency to coherence) remains nonzero and the belief dynamics still push the belief out of zero.
Even if the contribution of all triads of which $x$ is part is equal to $0$, the value of $x$ can change as a consequence of the social interactions with ``nonzero'' partners.

The belief dynamics can be made stochastic, for instance as following:
\begin{equation}\label{eq:prob_distribution}
    f(b_x^j(t),B^i(t)) \sim \mathcal{N} \left(\alpha b_{x}^{j}(t) - \beta\frac{\partial D^{i}(t)}{\partial b_x^{i}(t)},\sigma^2\right)
\end{equation}
where \(\mathcal{N}(\mu,\sigma^2)\) is a normal distribution with mean \(\mu\) and variance \(\sigma^2\), with \(\mu\) being the same as Eq.~\ref{eq:competition}.

A property of Eq.~\ref{eq:update_rule} is that an individual can change the strength of a belief even if they interact with another individual that attributes the same strength to that belief.  
This can be a consequence of the individual's predisposition to be coherent, or due to  stochasticity in Eq.~\ref{eq:prob_distribution}, or due to confirmation bias (which makes the belief stronger). 
Also, note that our formulation necessitates social influence for any belief update to occur.
Although it is possible to introduce various spontaneous belief dynamics, empirical observations show that attention to incoherencies is pivotal for their resolution ~\cite{van2021moral}. 
In contrast to conventional models like the bounded confidence model where the beliefs would remain the same when interacting with a same-belief partner~\cite{hegselmann2002opinion,deffuant2000mixing}, our model is more expressive and may capture a richer array of empirically known phenomena~\cite{bail2018exposure}. 

\section{Results}

Let us now discuss the emergent dynamics of the weighted belief network model. 
We begin with the simplest setup where the social network is a star graph and follow the evolution of the hub’s belief system in reaction to that of its neighbors (the leaf nodes). 
From this, we observe that, depending on the belief configuration, the resulting dynamics can be either simple or complex contagion-like in nature. 
We then approach the problem analytically using Markov state machines to verify the results obtained from simulations. 
Next, we move beyond the simplicity of the star graph and study the influence of network structure on contagion dynamics to offer more evidence for our findings. We first generate social networks using the Watts-Strogatz model~\cite{watts1998collective}.
This allows us to focus on the effect of clustering—a discriminant feature between simple and complex contagion—on the weighted belief dynamics. 
We then show that when a stable belief system is diffusing in a population of unstable beliefs, a random network structure is more beneficial for the contagion, concurring simple contagion dynamics. 
By contrast, when a stable belief system is diffusing in a population of another stable belief system, a clustered network is better than a random network, aligning with complex contagion dynamics~\cite{centola2010spread}.
This bolsters our claim that our model displays both simple and complex contagion dynamics. 
Last, we generate social networks using the stochastic block model~\cite{holland1983stochastic} and simulate the weighted belief dynamics on it. 
This choice allows us to focus on the dependence on the presence of community structure, another discriminant feature between simple and complex contagion.
Our setup shows the presence of optimal modularity behavior—a phenomenon observed only in complex contagion dynamics~\cite{nematzadeh2014optimal}(also see section S5)---further highlighting the weighted belief network model’s capacity to capture such dynamics.

\subsection{Emergence of simple and complex contagion}
\label{sec:spec_of_contagions}

\begin{figure*}
    \centering
    \includegraphics[width=1\textwidth, trim={0in 0.5in 0in 0.8in}]{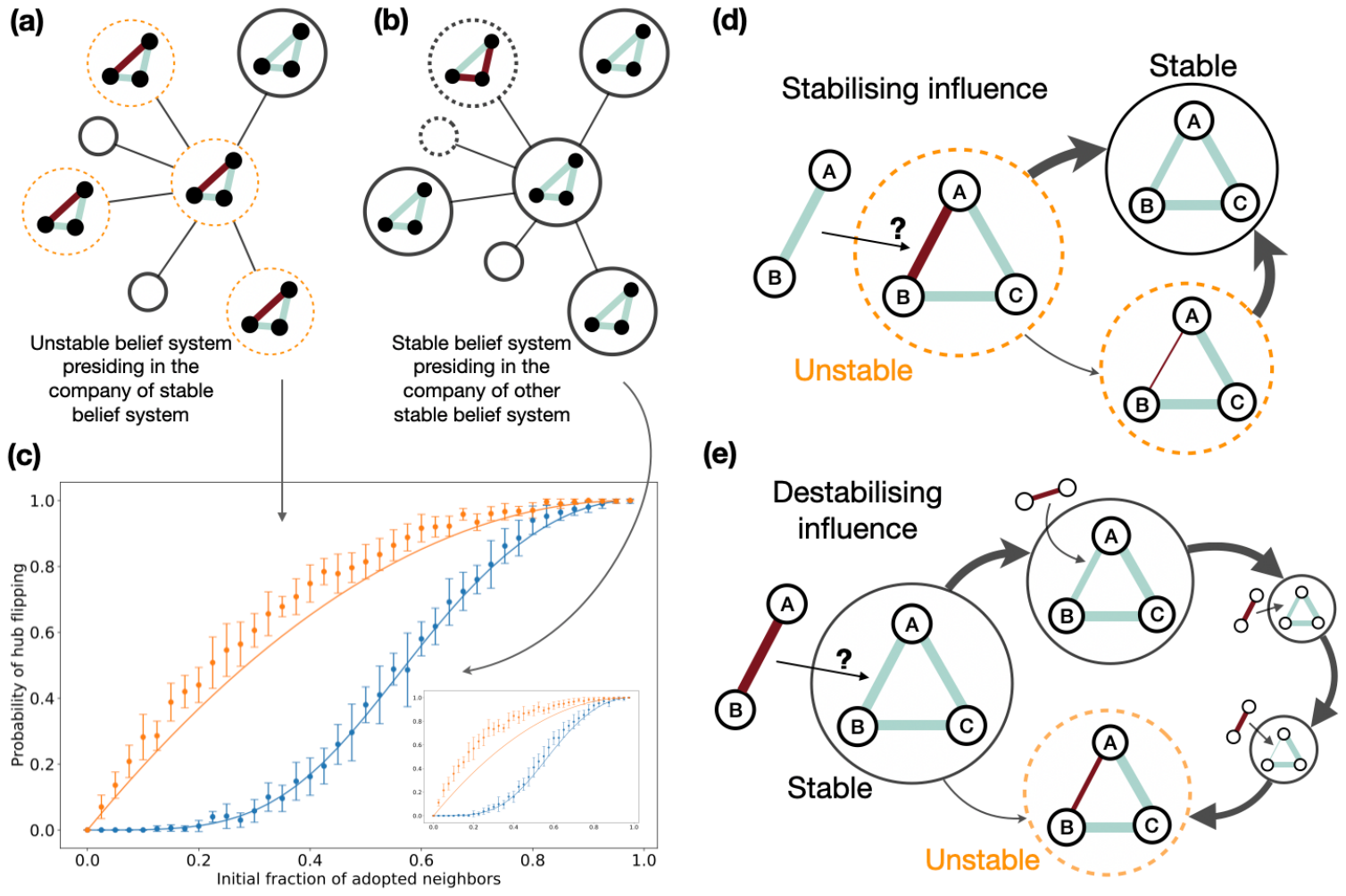}
    \caption{(a,b,c) Both simple and complex contagion dynamics emerge from the weighted belief network model. 
    (a) An unstable hub surrounded by a fixed fraction of stable neighbors. 
    (b) A stable hub surrounded by a fixed fraction of stable neighbors of a different kind. 
    (c) The probability that the hub changes to the new belief system exhibits the characteristic behaviors of simple (orange) and complex (blue) contagion  (\(N=40, M=39,\sigma=0.2,\alpha=1.5,\beta=1\)).
    The probability is calculated by running the simulation 50 times and calculating the proportion of times the hub “flipped.” Error bars, indicating SD, are then obtained by repeating this process 10 times. Dotted lines are results from numerical simulations and solid lines are results from the analytical calculations. 
    Inset: The numerical simulations carried out with the belief systems of neighbors of the hub which are initially similar to it are allowed to vary during the simulation. The robustness of results across different $\alpha$ and $\beta$ is shown in fig. S1 and across different $\sigma$ is shown in fig. S3. 
    (d,e) Intuition for the effect of social influence on an individual’s belief system.
    (d) An unstable belief system can easily be collapsed into a stable state with social influence. 
    (e) On the contrary, a stable individual’s belief system resists social influence that destabilizes it. Repeated exposures are necessary to push the individual into an unstable state.
    \label{fig:results_simplecomplex}}
\end{figure*}

Let us begin with a simplified case of a “star network” where everyone is connected exclusively to a single individual serving as a hub. Each individual has a fully connected belief network with three nodes representing the same notions between individuals (see Fig.~\ref{fig:results_simplecomplex}a,~b).

\paragraph{Scenario 1}Stabilising contagion behaves like a simple contagion (see Fig.~\ref{fig:results_simplecomplex}a,~d)   

At the start of the simulation, the hub has an unstable belief network with beliefs $\{-1, +1, +1\}$. 
A fraction of its neighbors has a stable belief network with weights $\{+1, +1, +1\}$
whereas the others have the same unstable belief network, i.e. $\{-1, +1, +1\}$.
During the course of the simulation, the belief systems of the leaf individuals are held fixed (``zealots") while that of the hub varies as the consequence of the pairwise interactions with its neighbors.
The fraction of stable individuals is the parameter varied in the simulations.

\paragraph{Scenario 2} Destabilising contagion behaves like a complex contagion system (see Fig.~\ref{fig:results_simplecomplex}b,~e)

At the start of the simulation, the hub has a stable belief network with weights $\{+1, +1, +1\}$. A fraction of its neighbors have a different stable belief network with weights $\{-1, -1, +1\}$ and 
the others have stable belief networks with edge weights identical to the hub, i.e. $\{+1, +1, +1\}$. 
During the course of the simulation, the belief system of the hub changes through pairwise interactions with its neighbors (zealots). The fraction of stable individuals having a stable belief network with weights
$\{-1, -1, +1\}$.  is the parameter varied in the simulations.

The only element of stochasticity here is the specific sequence of interactions and the noise in the belief update ($\sigma$ in Eq.~\ref{eq:prob_distribution}).
We consider the fraction of the simulations in which the belief system of the hub flips to the state of its discordant neighbors, as a function of their number. 
Results are shown in Fig.~\ref{fig:results_simplecomplex}c. 

Our results demonstrate that depending on the scenario—the context in which beliefs exist—considered, both simple and complex contagion dynamics can emerge from our model. 
When a stabilizing contagion stabilizes unstable belief systems, the change in belief polarity follows the simple contagion mechanism (scenario 1), as exemplified by the concave shape of the orange curve in Fig.~\ref{fig:results_simplecomplex}c.
When a new stable belief system competes with an already established stable belief system, the dynamics manifest itself as a complex contagion (scenario 2), characterized by the sigmoidal shape of the blue curve in Fig.~\ref{fig:results_simplecomplex}c.

These results also hold when the belief systems of the neighbors, which are initially similar to the hub, are allowed to vary, thus treating only the dissimilar neighbors as zealots (see inset, Fig.~\ref{fig:results_simplecomplex}).  
Our results confirm empirical observations suggesting that hard/costly changes in behavior (as those between two already stable belief systems) are typically driven by a complex contagion dynamics ~\cite{redlawsk2010affective}.

To summarize, depending on the current status of a belief network (stable versus unstable) and the nature of the social influence (stabilizing versus destabilizing), the social influence may behave like a simple contagion (stabilizing influence to unstable individuals; Fig.~\ref{fig:results_simplecomplex}d) or a complex contagion (destabilizing influence to stable individuals;  Fig.~\ref{fig:results_simplecomplex}e). 
Although a stable belief system resists destabilizing influence, repeated exposures can erode and eventually destabilize the belief system, which can then move into a different stable belief system.

\subsection{Analytical Approach}

The expected behavior of the model with the deterministic update rule (Eq.~\ref{eq:competition}) can be studied analytically. 
The evolution of the belief system can be described as a Markovian process over a finite and discrete set of states representing the belief system.
 
\begin{figure}[t]
\centering
    \includegraphics[width=0.5\textwidth, trim={1.5in 2in 1.5in 2in}]{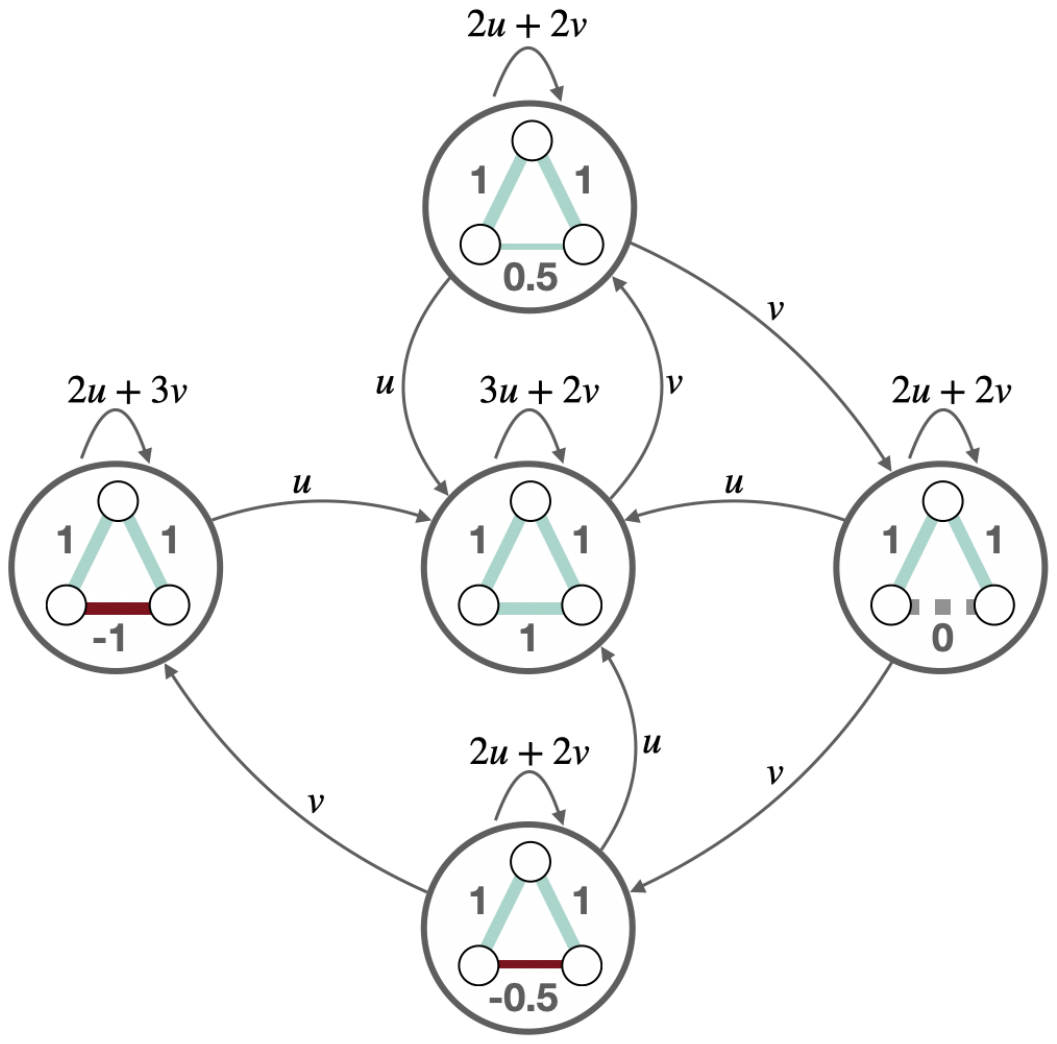}
    \caption{The state machine for Scenario 1 in the star graph setup with \(\alpha=1.5\) and \(\beta=1\). \(u (v)\) is proportional to the probability that the hub receives a belief from neighbors dissimilar (similar) to its initial state \label{fig:simple_statemachine}}
\end{figure}

The set of states and the transition probabilities for scenarios 1 and 2 are obtained using Eq.~\ref{eq:update_rule} and~\ref{eq:competition}. 
The parameters of the model are set as \(\alpha =1.5\) and \(\beta=1\) (from Fig.~\ref{fig:results_simplecomplex}). 
Given that we are using the deterministic update rule and the beliefs of the hub’s neighbors do not change, the set of states the hub can visit is relatively small for scenario 1 and is shown in Fig.~\ref{fig:simple_statemachine}. 
The respective transition matrix is given in Eq.~\ref{eq:simple_trasnition}. 

In the transition matrix $\pi$, the element $\pi_{xy}$ represents the probability of transition from state $y$ to $x$. 
$u$ is the probability that the hub receives a belief from any of its neighbors whose belief system is different from that of itself. 
If $m$ is the number of such neighbors (the x axis of Fig.~\ref{fig:results_simplecomplex}c) and $k$ is the total number of neighbors (degree of the hub), then \(u=\frac{m}{3k}\) where the \(3\) in the denominator is due to the fact that any one of the three beliefs in the dissimilar neighbor's triad can be chosen as the basis for the interaction. 
Similarly, \(v\) is the probability that the hub receives a belief from any of its neighbors whose belief system is the same as the hub's initial system. Therefore, \(v = \frac{k-m}{3k}\). 
Note, \(3(u+v)=1\), which is the sum of each column of the transition matrix.

\begin{equation}\label{eq:simple_trasnition}
    \pi = 
    \begin{blockarray}{cccccc}
        & $\{-1,1,1\}$ & $\{1,1,1\}$ &  $\{0.5,1,1\}$ & $\{0,1,1\}$ & $\{-0.5,1,1\}$ \\
      \begin{block}{c(ccccc)}
        $\{-1,1,1\}$ & 2u+3v & 0 & 0 & 0 & v \\
        $\{1,1,1\}$ & u & 3u+2v & u & u & u \\
        $\{0.5,1,1\}$ & 0 & v & 2u+2v & 0 & 0 \\
        $\{0,1,1\}$ & 0 & 0 & v & 2u+2v & 0  \\
        $\{-0.5,1,1\}$ & 0 & 0 & 0 & v & 2u+2v \\
      \end{block}
    \end{blockarray}
\end{equation}

The normalized eigenvector of the transition matrix associated with the eigenvalue $1$ gives the ($m$-dependent) stationary probability of the hub being in any state.

The probability of the hub flipping its initial unstable belief system, $\{-1,+1,+1\}$, to a stable belief system (that of its neighbors who are different from the hub) is then evaluated as the sum of the probabilities of the hub being either in state $\{1,1,1\}$ or $\{0.5,1,1\}$ at stationarity. 
This results in the orange solid curve in Fig.~\ref{fig:results_simplecomplex}c, which closely approximates the orange dotted curve obtained from numerical simulations of the stochastic version (Eq.~\ref{eq:prob_distribution}).
The Markov process for Scenario 2 is obtained following a similar strategy. This results in a larger number of states (20) and can be approached similarly to Scenario 1 (see section S3).
The probability of the hub flipping its initial stable belief system, $\{-1,-1,+1\}$, to a stable belief system of a different kind (the belief system of the neighbors who are different from the hub) is then evaluated by the sum of the probabilities of the hub being either in state $\{1,1,1\}$ or $\{1,0.5,1\}$ or $\{0.5,1,1\}$ at stationarity.
This results in the blue solid curve in Fig.~\ref{fig:results_simplecomplex}c which concurs with the solid curve obtained from numerical simulations of the stochastic version (Eq.~\ref{eq:prob_distribution}).

\subsection{Influence of the network structure on the contagion dynamics
}

The structure of the underlying social network plays an important role in the diffusion of information. 
A network with many clusters and a large diameter will be less effective for simple contagion than its random counterparts, which—at the expense of locally redundant ties—provides shortcuts that connect remote regions.
By contrast, the clustered network can enhance the spreading of complex contagion by facilitating social reinforcement locally~\cite{centola2010spread}.

We expect, therefore, that by setting the parameters of our model in the ``complex contagion'' regime, it will spread more easily in a highly clustered network, whereas the opposite would happen in the ``simple contagion'' regime.

We test this hypothesis by simulating our model on a highly clustered network [Watts-Strogatz network~\cite{watts1998collective} with rewiring probability, $P = 0$] and a random network (Watts-Strogatz network with rewiring probability $P = 1.0$)—the two network settings used by Centola~\cite{centola2010spread}.

\paragraph{Scenario 1}

Does simple contagion spread faster on a random network than on a clustered network?

\begin{figure}
\centering
    \includegraphics[width=0.8\textwidth, trim={1.5in 1in 1.5in 2in}]{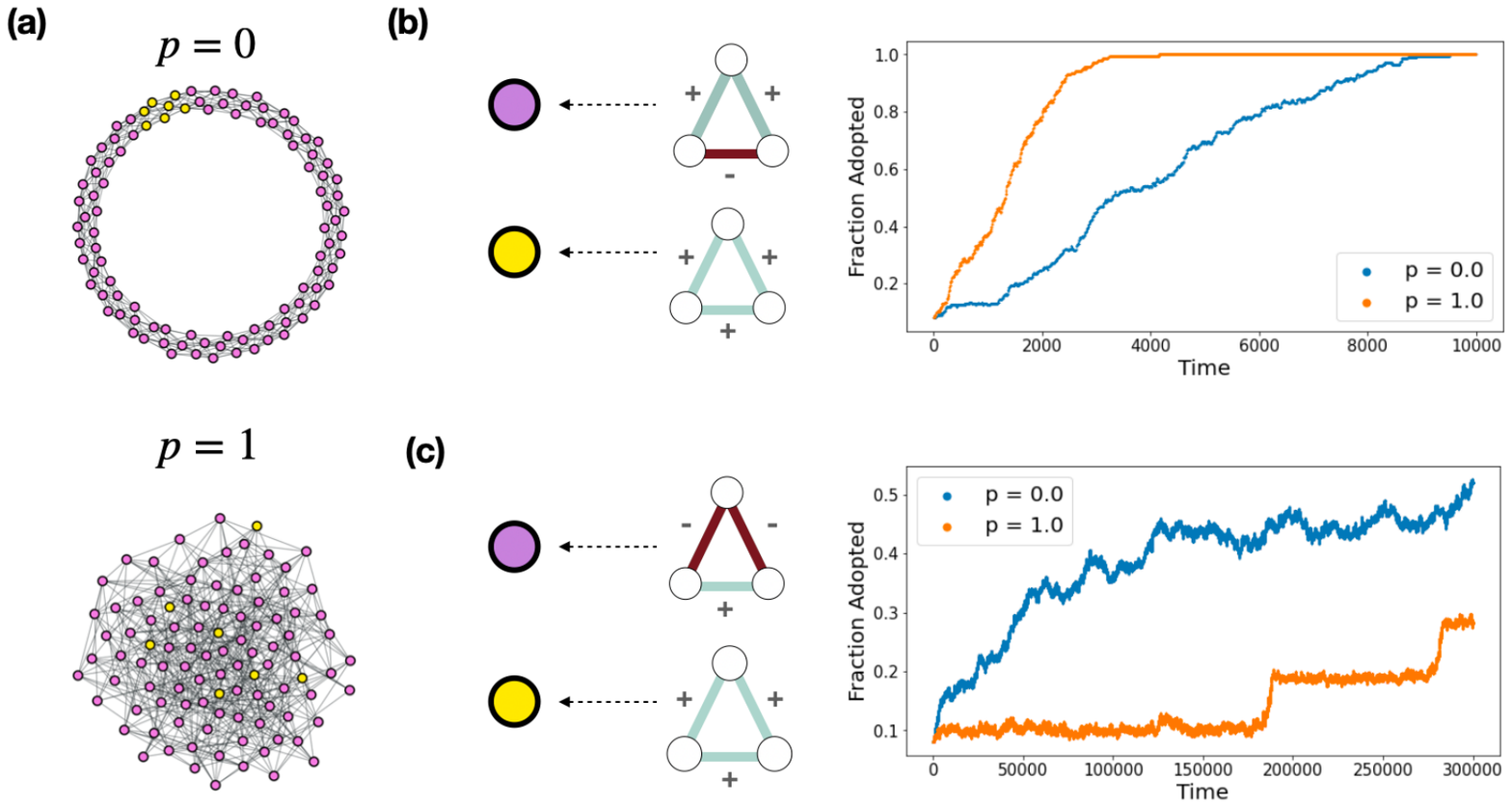}
    \caption{The impact of network structure on the belief cascade.
    We run the two scenarios (“simple” versus “complex”) on two networks (clustered large world versus random small world). 
    As predicted by the theory of simple and complex contagion, we observe that a new stabilizing belief (simple contagion) spreads better in a random small world than in a clustered large world, whereas a new stable belief system that competes with an existing stable belief system (complex contagion) spreads better in a clustered large world rather than a random small world. 
    Yellow nodes represent the seeds whose belief systems do not change during the course of the simulation (zealots), and purple represents the rest of the population.
    (a) Watts-Strogatz model with rewiring probability $P = 0$ results in a clustered large world, whereas $P = 1$ results in a random small world network. 
    (b) Scenario 1: Simple contagion spreads faster in a random network ($P = 1$) than in a clustered network ($ P = 0)$. 
    (c) Scenario 2: Complex contagion spreads faster on a clustered network ($P = 0$) than on a random network ($P = 1$).
    $\alpha=2,\beta=1,N=100,\bar{k}=10,\sigma=0.2,\rho_0=0.08$. 
     Curves are obtained from averaging 10 ensembles. See the section S4 for varying $\rho_0$. \label{fig:watts_strogatz}}
\end{figure}

As described above, we use the WS network with $P = 0$ and $P = 1.$
Everyone’s belief system is initially set in the unstable configuration $\{-1, +1, +1\}$. 
We then choose a fraction of individuals ($\rho_0$) and set their belief to the stable configuration $\{+1, +1, +1\}$. 
The belief systems of this seed set are held fixed during the simulation. 
Similar to the initialization in Centola~\cite{centola2010spread}, seed nodes are concentrated for the $P = 0$ network while seeds are randomly chosen in the $P = 1$ case (see Fig.~\ref{fig:watts_strogatz}a). 
The results shown in Fig.~\ref{fig:watts_strogatz}b demonstrate that, as expected, the stable-spreading-into-unstable setting produces results consistent with simple contagion: spreading is more effective on a random network than on a highly-clustered one.

\paragraph{Scenario 2}

Does complex contagion spread faster on a clustered network than on a random network?

Here, with the same two-network setup, we have two competing, stable belief systems. 
Every belief system is initially set to be $\{-1, -1, +1\}$ and then the belief system of the seed set ($\rho_0$) is set to be $\{+1, +1, +1\}$ and held fixed during the course of the simulation. 
In all other aspects, the setting is analogous to that described in scenario 1. 
This scenario produces richer dynamics (see section S4.2).
When $\rho_0$ is small, the highly clustered network spreads the contagion better, replicating the phenomenon observed by Centola (see Fig.~\ref{fig:watts_strogatz}c)~\cite{centola2010spread}. 
However, as we increase $\rho_0$, the contagion spreads better on the small-world network (see section S4.2). 
Although this was not reported in Centola’s experiment, it is a reasonable outcome. 
As we increase the number of seeds, the chance they induce multiple reinforcements also goes up. 
When the number of seeds is high and they are distributed across the whole network, they can simultaneously induce multiple reinforcements across the whole network, producing a rapid global cascade.

\subsection{Optimal Modularity}

The community structure of a social network plays an important role in the diffusion of information. 
For instance, a strong community structure and ``thin'' bridges can act as a bottleneck for contagion ~\cite{centola2007complex}. 
At the same time, a strong community structure facilitates social reinforcement and enhances local spreading in scenarios characterized by complex contagion~\cite{centola2010spread}. 
A recent study explored systematically how the modular structure of networks influences information diffusion, showing that there exists a region of optimal modularity where community structure counterintuitively enhances rather than hinders global diffusion of complex contagion~\cite{nematzadeh2014optimal}. 
We expect the same behavior also in our model when it is characterized by complex contagion dynamics. 

\begin{figure*}
\centering
    \includegraphics[width=0.7\textwidth, trim={1.5in 1in 1.5in 1in}]{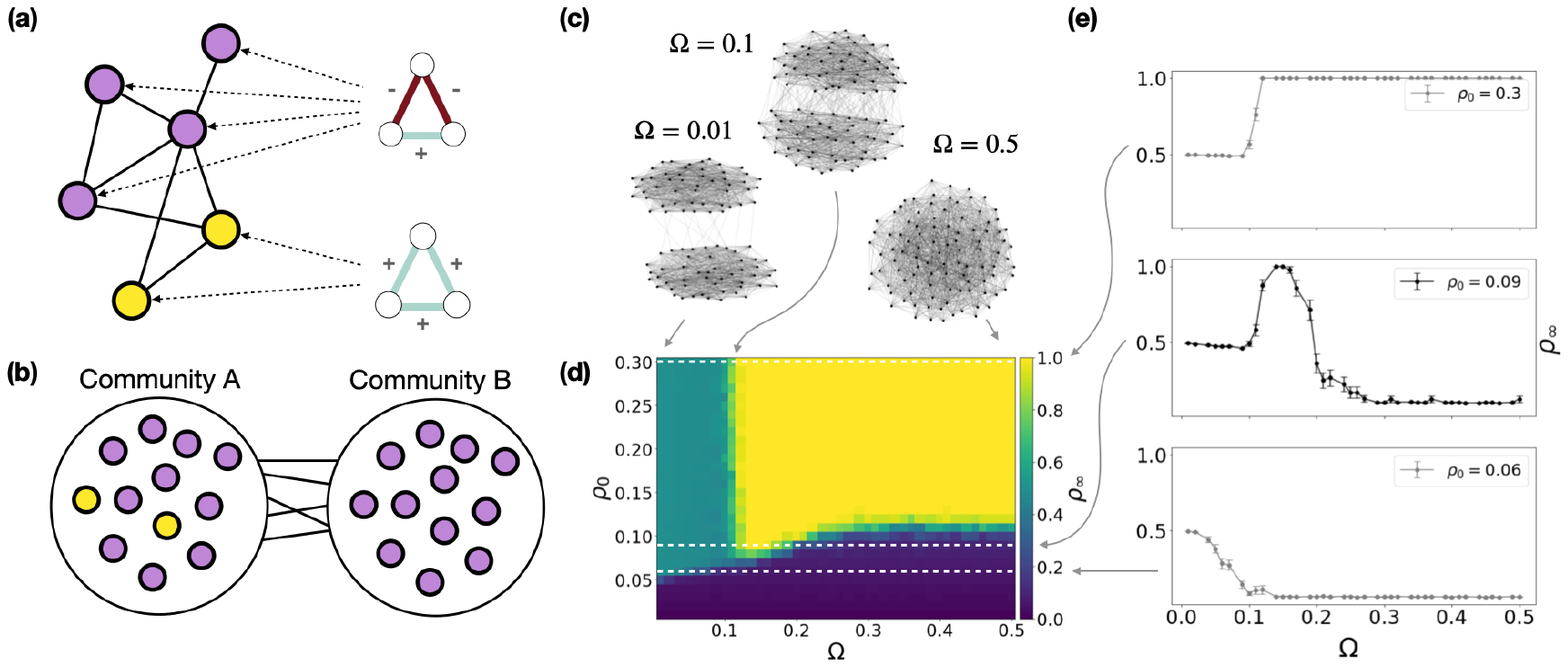}
    \caption{The case of complex contagion (two competing stable belief systems) also exhibits the optimal modularity behavior. (a,b) An illustration of the initial condition of the simulation. The population (purple nodes) has a stable belief system and the zealots (yellow nodes), constrained to a single community,  have a stable belief system of a different kind. (c) The mixing parameter \(\Omega\) determines the strength of community structure in the network. (d) The phase diagram exhibits optimal modularity with three phases: no diffusion (dark blue), local diffusion in the seed community (green), and global diffusion (yellow). (e) Cross sections of the phase diagram. Error bars indicate standard error. The following parameters are used: \(N=100, M=1500,\sigma=0.2,\alpha=2,\beta=1,40\) ensembles. \label{fig:results_optimalmodularity}}
\end{figure*}

Fig.~\ref{fig:results_optimalmodularity} demonstrates this is indeed the case. 
As shown in Fig.~\ref{fig:results_optimalmodularity}c we consider an ensemble of social networks having $N$ nodes and $M$ edges generated using the stochastic block model~\cite{holland1983stochastic} with nodes equally distributed in two communities (blocks). We introduce a parameter $\Omega$ to control the strength of the two communities. 
The probability for nodes in the same community to be connected to each other is given by $\frac{4(1-\Omega M}{N(N-2)}$ and and that of nodes in different communities is $\frac{4\Omega M}{N(N-2)}$.
A large value of $\Omega$ results in more links between the two communities and thus a weaker community structure.

Next, similar to the setup of belief systems in the case of complex contagion in Fig~\ref{fig:results_simplecomplex}b, a fraction \(\rho_0\) of nodes in the social network, belonging to the same community are initialised with stable belief networks of the kind $\{+1,+1,+1\}$. 
The belief networks of these individuals are held fixed throughout the simulation (zealots). 
The remaining set of nodes are initialised with stable belief networks of a different kind, $\{-1,-1,+1\}$. 
An example of this setup is shown in Fig.~\ref{fig:results_optimalmodularity}a, b.

Once \(\alpha\) and \(\beta\) have been fixed, we compute \(\rho_\infty\), the fraction of nodes that have adopted the zealots' stable belief system at stationarity via numerical simulations. 
For small values of \(\rho_0\) \((\rho_0 < 0.06)\), the new stable belief system is not adopted by the nodes with an already stable belief system and contagion essentially fails to propagate regardless of \(\Omega\) (see Fig.~\ref{fig:results_optimalmodularity}d). 
At, or just above $\rho_0 = 0.06$, all the nodes in the originating community adopt the new belief system if \(\Omega\) is sufficiently small. 
However, when a critical value of \(\Omega\) is exceeded, the intra-community connectivity becomes insufficient to spread the new stable belief system to the whole originating community 
(Fig.~\ref{fig:results_optimalmodularity}e bottom).

A larger value of \(\rho_0\) \((\rho_0 = 0.09\)) finally allows for global diffusion of the new stable belief system in the presence of a sufficient number of bridges between the two communities.
However, increasing the inter-community connectivity at the expense of intra-community connectivity (increasing \(\Omega\)), does not lead to the new stable belief systems being adopted in the originating community and therefore it cannot be transmitted over the entire network, despite the increased number of inter-community links. 
This is reflected in the intermediate range of community strength that allows global adoption of the new stable belief system (Fig.~\ref{fig:results_optimalmodularity}e middle). 
The presence of a regime of optimal network modularity, where the adoption of the new stable belief system is maximized, is a phenomenon unique to complex contagion (see section S5 for our mathematical argument that this phenomenon does not appear in simple contagion) and further buttress the idea that our weighted belief model can exhibit both simple and complex contagion dynamics depending on the stability of belief systems and the type of the incoming belief. 
When \(\rho_0\) becomes even larger, increasing \(\Omega\) doesn't block the local spreading anymore, and thus the global diffusion always happens as long as the network has enough bridges (Fig.~\ref{fig:results_optimalmodularity}e top).

\section{Discussion and Limitations}

Here, we demonstrate that both simple and complex contagion can emerge from a simple model of the weighted belief network that is rooted in the fundamental cognitive predisposition to achieve internal coherence. 
The interactions between the existing beliefs of an individual, when combined with the propensity to achieve internal coherence, provide resistance to incompatible beliefs coming from the individual’s social neighborhood. 
Such resistance can be overcome by social reinforcement, leading to complex contagion dynamics. 
Our model elucidates how resistance may act as the key mechanism that produces the dynamics of complex contagion.
An original contribution of our work is that simple and complex contagion dynamics emerge from our model depending on the context—the network of social interaction and the specific distribution of beliefs across the population.

The belief system in this work has been operationalized as a network where the nodes are concepts, entities, or notions, and the edges are beliefs. 
In such an operationalization, the nodes of the belief system are free of attributes. 
This differs from some recent work which considered beliefs as nodes and the interdependencies between them as edges~\cite{galesic2021integrating}. 
In such a case, both nodes and edges have attributes to them. 
Comparing the two operationalizations would be an interesting direction for future research.

This work considers the simplest form of a system of interacting beliefs (a triad) and does not distinguish between different types of nodes and edges in a belief system. 
The structure of the belief triad is static but exploring the effect of dynamical belief networks on contagion dynamics can be an interesting direction for future research. It also treats social interactions to be unweighted and undirected and occur on a static network, with an equal rate of interaction across all pairs. 
In addition, an assumption made in Eq.~\ref{eq:update_rule} is that the belief of the focal individual being updated to decrease dissonance is the same as the one being socially influenced by its neighbors.
In practice, however, an individual might change a different belief, say by, to reduce dissonance in reaction to the incoming belief, say bx, from its social neighbor~\cite{dalege2022using}. 
Incorporating the complexity of human social interactions and examining their implications on large-scale social phenomena will be an interesting direction for future work. It will be also fruitful to directly evaluate both the microscopic and macroscopic social dynamics that are predicted by our model.

Eventually, empirically validated and calibrated models of contagion may facilitate our understanding of how, where, and why misinformation permeates and spreads. 
However, we acknowledge that such a model of belief dynamics may not be necessary to explain, for instance, why rumors follow simple contagion, and protesting follows complex contagion---the risk involved with each action (translated into an adoption threshold) may be enough and accounting for prior beliefs may not be necessary. 
Despite numerous limitations, we believe that our demonstration of rich dynamics from belief interactions provides a strong rationale for developing and using social contagion models that are more firmly grounded in cognitive processes.

\section*{Acknowledgements}

We thank F. Menczer, J. Bollen, and R. Goldstone, the members of the cognitive science department at Indiana University, for helpful conversations. 
We also thank a reviewer for the idea of conducting Centola’s experiment. 
We acknowledge the computing resources at the Luddy School, Indiana University, which were key for the simulations performed in this paper.

\section*{Data and materials availability}

Code is available in \url{10.5281/zenodo.8370688} (or \url{https://github.com/rachithaiyappa/beliefnet}). 
All other data needed to evaluate the conclusions in this paper are present in the paper and/or the Supplementary Materials.

\printbibliography
\newpage
\begin{titlepage}
   \begin{center}
       \vspace*{1cm}
        \Large Supplementary Materials for\\
        \vspace*{2cm}
       \textbf{Emergence of Simple and Complex Contagion Dynamics from Weighted Belief Networks}

       \vspace{0.5cm}
        Rachith Aiyappa,$^{1*}$ Alessandro Flammini,$^1$ Yong-Yeol Ahn$^1$
            
       \vspace{1cm}

       \large{$^1$Center for Complex Networks and Systems \\
       Luddy School of Informatics, Computing, and Engineering \\
       Indiana University Bloomington\\ Bloomington, Indiana, USA, 47408.\\
       \vspace{1cm}
	$^*$Corresponding author: racball@iu.edu}
            
   \end{center}
\end{titlepage}
\clearpage

\setcounter{section}{0}
\renewcommand*{\theHsection}{chX.\the\value{section}}
\renewcommand\thesection{S\arabic{section}}

\setcounter{figure}{0}
\renewcommand*{\theHfigure}{ chX.\the\value{figure}}
\renewcommand\thefigure{S\arabic{figure}}

\setcounter{table}{0}
\renewcommand*{\theHtable}{chX.\the\value{table}}
\renewcommand\thetable{S\arabic{table}}

\setcounter{equation}{0}
\renewcommand*{\theHequation}{chX.\the\value{equation}}
\renewcommand\theequation{S\arabic{equation}}

\section{Varying $\alpha$ and $\beta$ }

\subsection{Star Graph Setting}
\begin{figure}[!htb]
    \hspace*{-1.5cm}
    \includegraphics[width=1.2\textwidth, trim={0 1in 0 1in}]{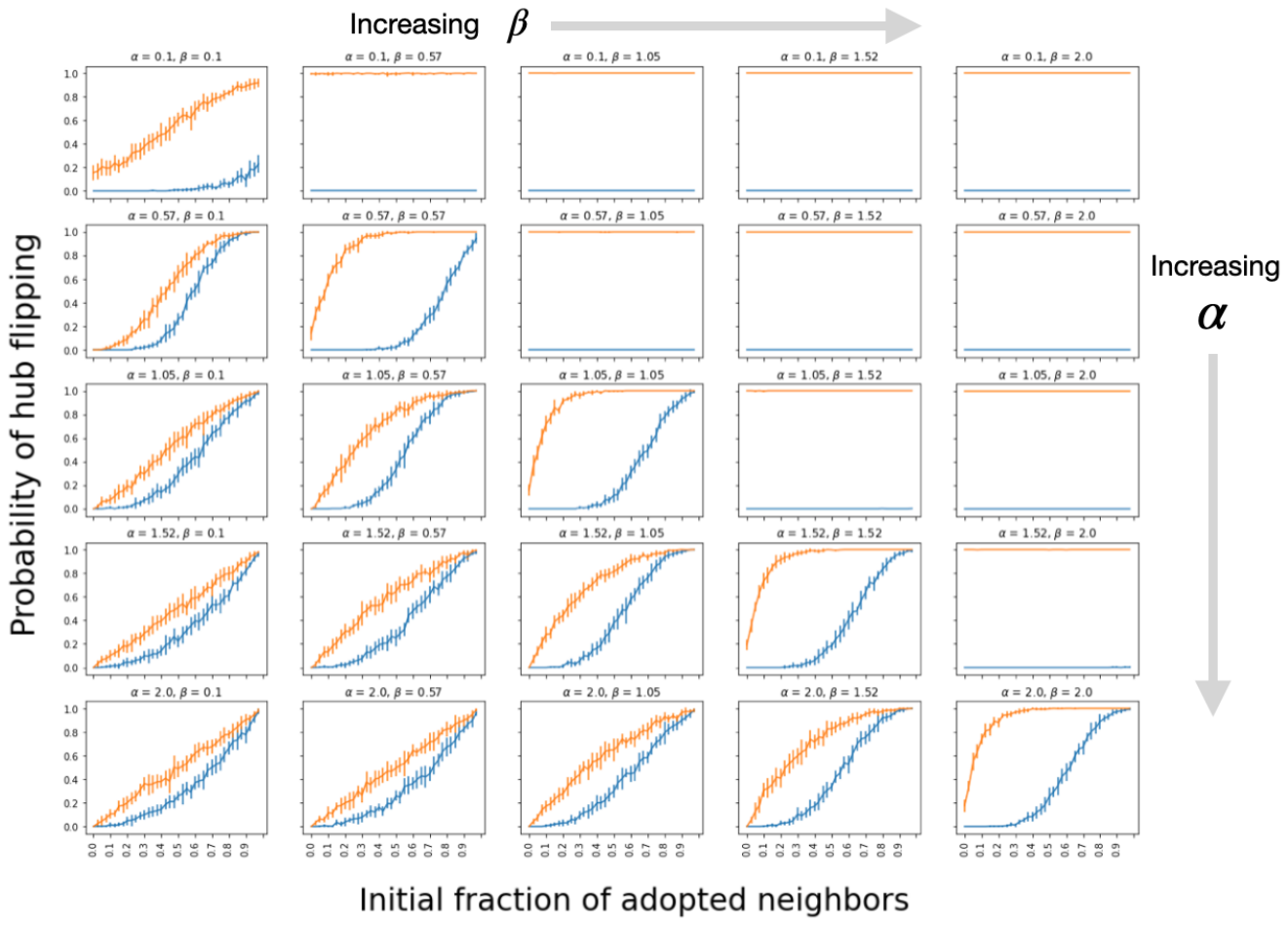}
    \caption{\textbf{The probability that the hub changes to the new belief system exhibits the characteristic behaviors of simple (orange) and complex (blue) contagion for a range of model parameters---$\alpha$ and $\beta$}. The probability is calculated by running the simulation 50 times and calculating the proportion of times the hub `flipped.' Error bars, indicating standard deviation, are then obtained by repeating this process 10 times. \(N=40, M=39,\sigma=0.2\).}\label{fig:varying_alphabeta}
\end{figure}
\clearpage

\subsection{Stochastic Block Model Setting}
\begin{figure}[!htb]
    \includegraphics[width=1\textwidth, trim={0 0in 0 0in}]{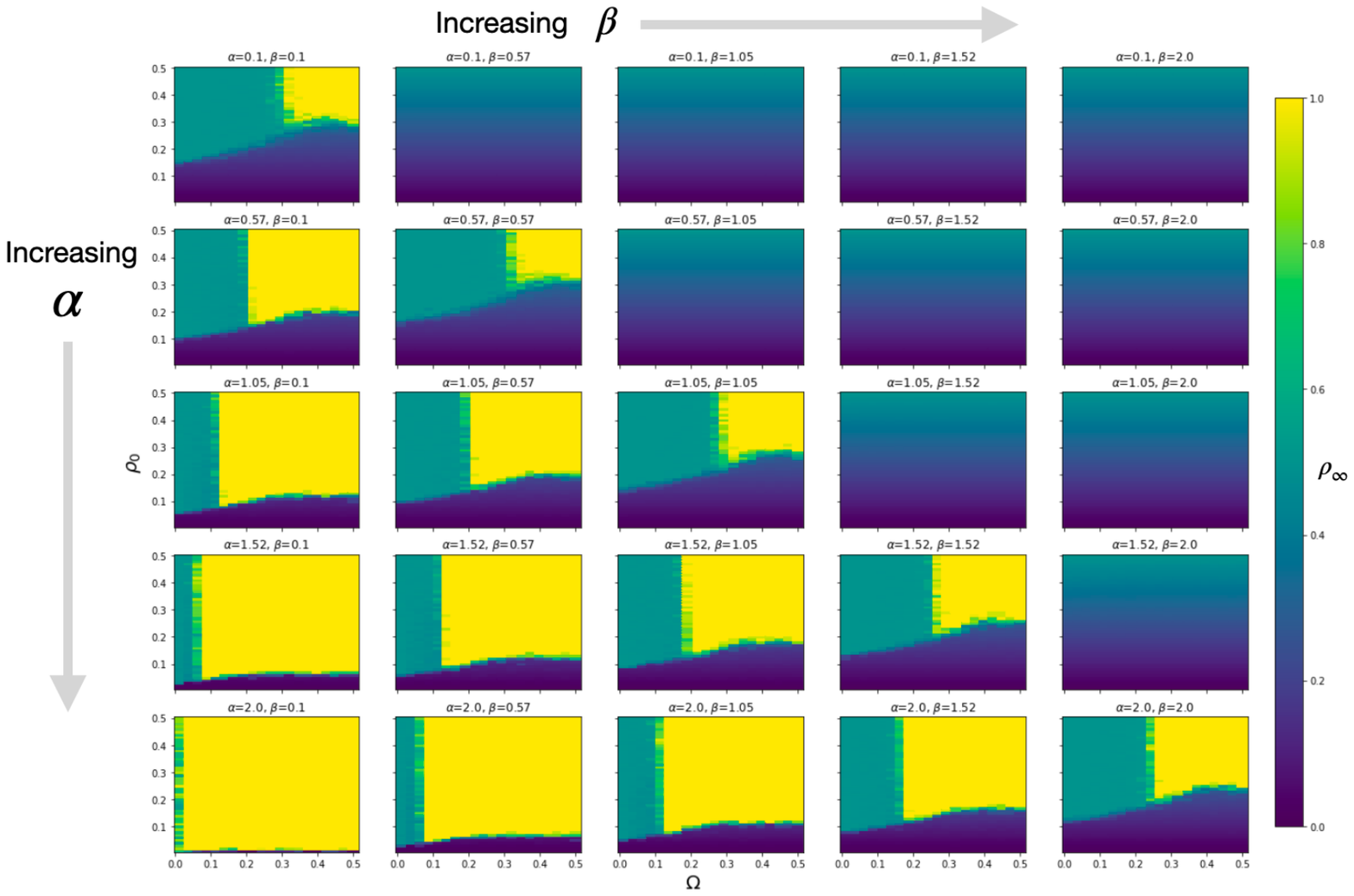}
    \caption{\textbf{The presence of optimal modularity behavior for different values of model parameters---$\alpha$, and $\beta$}. No diffusion (dark blue), local diffusion in the seed community (green), and global diffusion (yellow). $N=100, M=1500, \sigma=0.2, 10 $ ensembles. }\label{fig:om_varying_alphabeta}
\end{figure}
\clearpage

\section{Varying $\sigma$}
\subsection{Star Graph Setting}
\begin{figure}[!htb]
    \hspace*{-1.5cm}
    \includegraphics[width=1.2\textwidth, trim={0 2.5in 0 3in}]{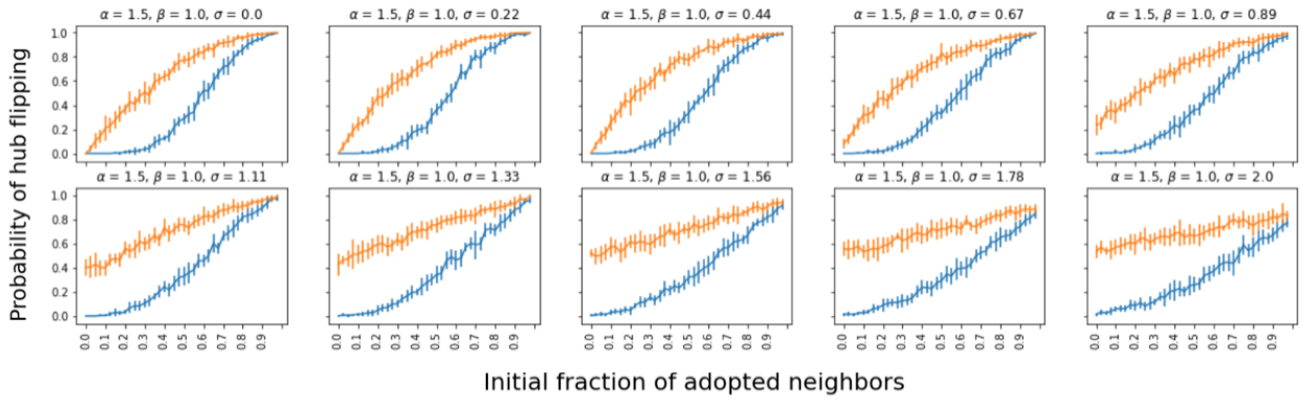}
    \caption{\textbf{The probability that the hub changes to the new belief system exhibits the characteristic behaviors of simple (orange) and complex (blue) contagion for different values of $\sigma$}. The probability is calculated by running the simulation 50 times and calculating the proportion of times the hub `flipped.' Error bars, indicating standard deviation, are then obtained by repeating this process 10 times. \(N=40, M=39\)}\label{fig:varying_sigma}
\end{figure}
\clearpage

\section{Transition probabilities for complex contagion}

Tab.~\ref{tab:states_scenario2} assigns an ID to each state in scenario 2 described in the main paper (complex contagion setting) with the deterministic update rule with $\alpha=1.5,\beta=1$. Tab.~\ref{tab:scenario2_transitions} shows the transition probabilities between states from which the transition matrix and normalized eigenvector corresponding to eigenvalue 1 can be obtained. 

\begin{table}[htbp]
    \centering
    \begin{tabular}{|c|c|}
        \hline
        \textbf{ID} & \textbf{Belief Network State} \\
        \hline
        0 & $\{-1, -1, 1\}$ \\
        1 & $\{-1, -0.5, 1\}$ \\
        2 & $\{-0.5, -1, 1\}$ \\
        3 & $\{-0.5, 0, 1\}$ \\
        4 & $\{-0.5, 1, 1\}$ \\
        5 & $\{-1, 0, 1\}$ \\
        6 & $\{0, -1, 1\}$ \\
        7 & $\{0.5, -1, 1\}$ \\
        8 & $\{-1, 0.5, 1\}$ \\
        9 & $\{-1, 1, 1\}$ \\
        10 & $\{0, 0.5, 1\}$ \\
        11 & $\{0, 1, 1\}$ \\
        12 & $\{0, -0.5, 1\}$ \\
        13 & $\{1, 0.5, 1\}$ \\
        14 & $\{1, 1, 1\}$ \\
        15 & $\{1, 0, 1\}$ \\
        16 & $\{0.5, 1, 1\}$ \\
        17 & $\{0.5, 0, 1\}$ \\
        18 & $\{1, -0.5, 1\}$ \\
        19 & $\{1, -1, 1\}$ \\
        \hline
    \end{tabular}
    \caption{\textbf{All the states the hub can transition to in scenario 2 of the star graph setting}. Each state is mapped to a  unique ID. }
    \label{tab:states_scenario2}
\end{table}

\begin{table}[htbp]
    \centering
    \begin{minipage}{0.2\linewidth}
        \begin{tabular}{|c|c|c|}
            \hline
            \textbf{SID} & \textbf{DID} & \textbf{TP} \\
            \hline
            0 & 0 & $u + 3v$ \\
            0 & 1 & $u$ \\
            0 & 2 & $u$ \\
            1 & 0 & $v$ \\
            1 & 1 & $u + 2v$ \\
            1 & 5 & $u$ \\
            1 & 12 & $u$ \\
            2 & 0 & $v$ \\
            2 & 2 & $u + 2v$ \\
            2 & 3 & $u$ \\
            2 & 6 & $u$ \\
            3 & 2 & $v$ \\
            3 & 3 & $u + v$ \\
            3 & 4 & $u$ \\
            3 & 5 & $v$ \\
            3 & 15 & $u$ \\
            4 & 2 & $v$ \\
            4 & 4 & $2u + v$ \\
            4 & 9 & $v$ \\
            4 & 14 & $u$ \\
        \hline
        \end{tabular}
    \end{minipage}
    \hspace{0.05\linewidth} 
    \begin{minipage}{0.2\linewidth}
        \begin{tabular}{|c|c|c|}
            \hline
            \textbf{SID} & \textbf{DID} & \textbf{TP} \\
            \hline
            5 & 0 & $v$ \\
            5 & 5 & $2v + u$ \\
            5 & 8 & $u$ \\
            5 & 17 & $u$ \\
            6 & 0 & $v$ \\
            6 & 6 & $u + 2v$ \\
            6 & 7 & $u$ \\
            6 & 9 & $u$ \\
            7 & 0 & $v$ \\
            7 & 6 & $u + 2v$ \\
            7 & 16 & $u$ \\
            7 & 19 & $u$ \\
            8 & 0 & $v$ \\
            8 & 8 & $u + 2v$ \\
            8 & 9 & $u$ \\
            8 & 14 & $u$ \\
            9 & 0 & $v$ \\
            9 & 9 & $2u + 2v$ \\
            9 & 14 & $u$ \\
            10 & 6 & $v$ \\
        \hline
        \end{tabular}
    \end{minipage}
    \hspace{0.05\linewidth} 
    \begin{minipage}{0.2\linewidth}
        \begin{tabular}{|c|c|c|}
            \hline
            \textbf{SID} & \textbf{DID} & \textbf{TP} \\
            \hline
            10 & 8 & $u$ \\
            10 & 10 & $u + v$ \\
            10 & 11 & $u$ \\
            10 & 13 & $u$ \\
            11 & 4 & $v$ \\
            11 & 11 & $2u + v$ \\
            11 & 12 & $v$ \\
            11 & 14 & $u$ \\
            12 & 1 & $v$ \\
            12 & 6 & $v$ \\
            12 & 11 & $u$ \\
            12 & 12 & $u + v$ \\
            12 & 18 & $u$ \\
            13 & 10 & $v$ \\
            13 & 13 & $2u + v$ \\
            13 & 14 & $u$ \\
            13 & 15 & $v$ \\
            14 & 13 & $v$ \\
            14 & 14 & $3u + v$ \\
            14 & 16 & $v$ \\
        \hline
        \end{tabular}
    \end{minipage}
    \hspace{0.05\linewidth} 
    \begin{minipage}{0.2\linewidth}
        \begin{tabular}{|c|c|c|}
            \hline
            \textbf{SID} & \textbf{DID} & \textbf{TP} \\
            \hline
            15 & 3 & $v$ \\
            15 & 14 & $u$ \\
            15 & 15 & $2u + v$ \\
            15 & 18 & $v$ \\
            16 & 11 & $v$ \\
            16 & 14 & $u$ \\
            16 & 16 & $2u + v$ \\
            16 & 17 & $v$ \\
            17 & 5 & $v$ \\
            17 & 7 & $u$ \\
            17 & 14 & $u$ \\
            17 & 15 & $u$ \\
            17 & 16 & $u + v$ \\
            18 & 1 & $v$ \\
            18 & 14 & $u$ \\
            18 & 18 & $2u + v$ \\
            18 & 19 & $v$ \\
            19 & 0 & $v$ \\
            19 & 14 & $u$ \\
            19 & 19 & $2u + 2v$ \\
        \hline
        \end{tabular}
    \end{minipage}
    \caption{\textbf{Transition Probabilities (TIP) from Source State (SID) to Destination State (DID) for Scenario 2 in the star graph setting}. Missing SID-DID combinations have a TP of 0.}
    \label{tab:scenario2_transitions}
\end{table}
\clearpage

\section{Influence of network structure on belief cascade}
\subsection{Scenario 1: Simple contagion setting}
\begin{figure}[!htb]
    \includegraphics[width=1\textwidth, trim={0 1.5in 0 0in}]{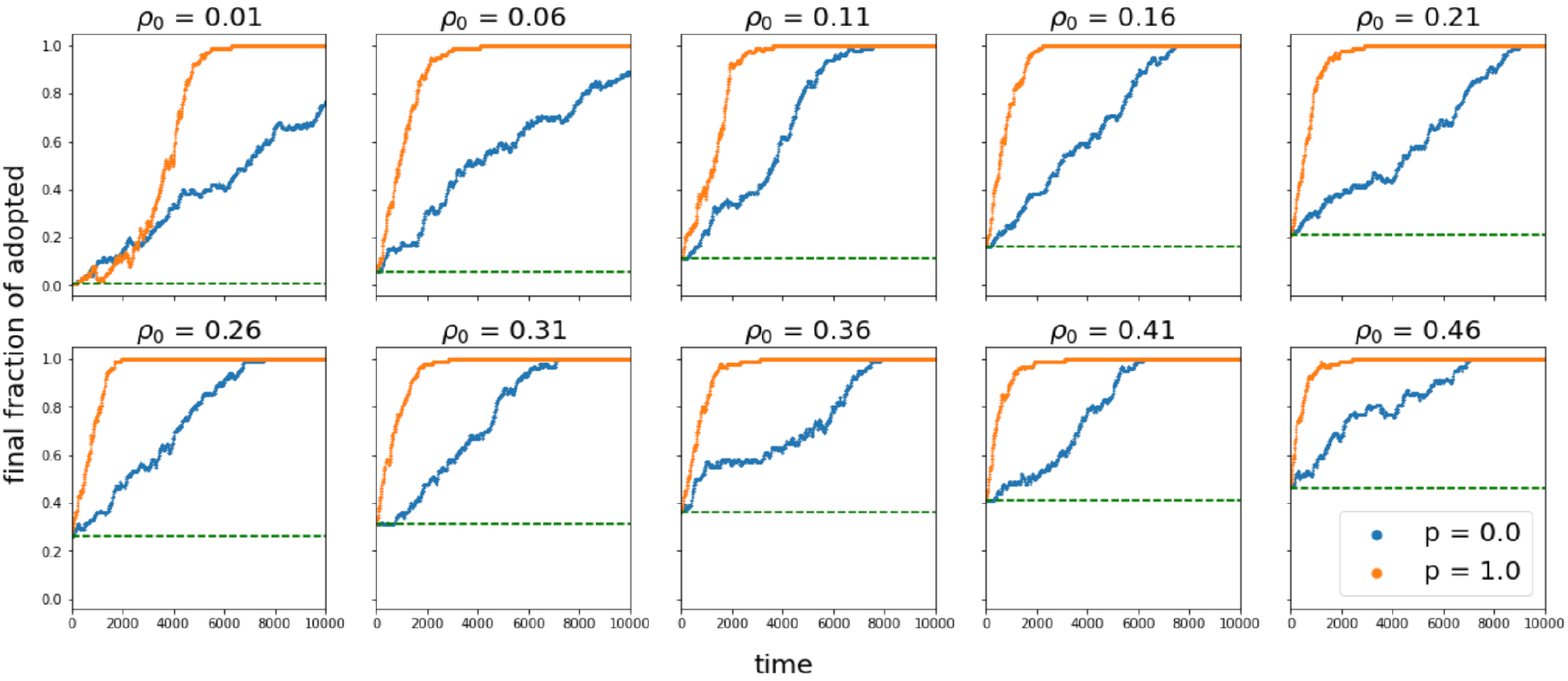}
    \caption{\textbf{The time evolution of the fraction of stable individuals of the type $\{+1,+1,+1\}$ in Scenario 1 of the Watts-Strogatz network setting.} We see that a random network structure (Watts Strogatz with rewiring probability $p=1$, orange curve) facilitates the contagion more than a highly clustered network structure (Watts Strogatz with rewiring probability $p=0$, blue curve). The horizontal green line indicates $\rho_0$. N=100, $\bar{k}=10, \sigma=0.2, 2$ ensembles.}\label{fig:wattstrogatz_simple}
\end{figure}
\clearpage
\subsection{Scenario 2: Complex Contagion setting}
\subsubsection{Average degree = 10}
\begin{figure}[!htb]
    \includegraphics[width=1\textwidth, trim={0 1in 0 0in}]{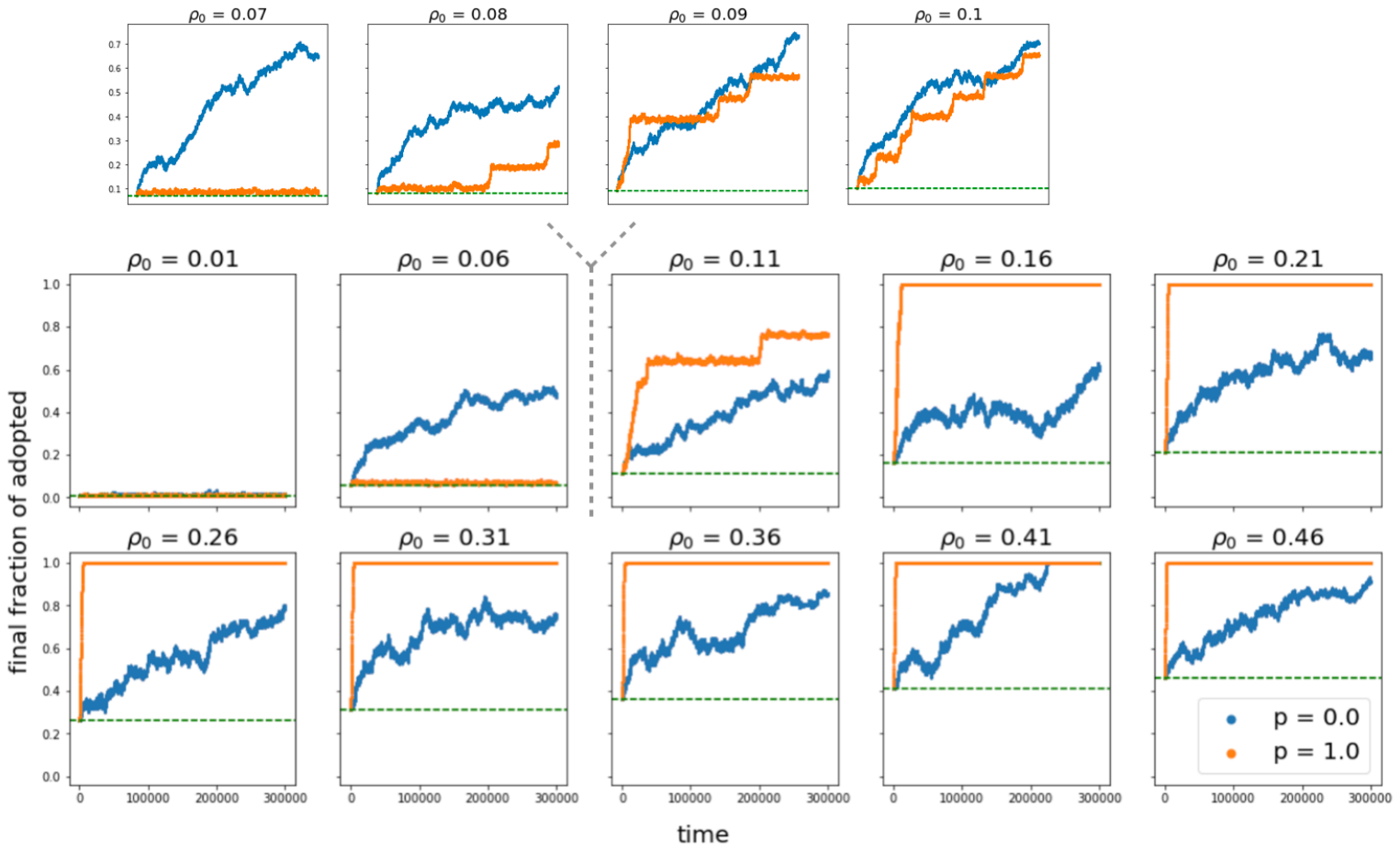}
    \caption{\textbf{The time evolution of the fraction of stable individuals of the type $\{+1,+1,+1\}$ in Scenario 2 of the Watts-Strogatz network setting.} We see a rich range of behaviors. In case of $\rho_0=0.06$, we see that a clustered network structure (Watts Strogatz with rewiring probability $p=0$, blue curve) facilitates the contagion more than a random network structure (Watts Strogatz with rewiring probability $p=1$, orange curve) while in other cases it appears that a random network structure facilitates the contagion more than a highly clustered network structure. The horizontal green line indicates $\rho_0$. N=100, $\bar{k}=10, \sigma=0.2$. Curves are obtained from averaging $3$ ensembles for all subplots except those between $\rho_0=0.06, \rho=0.11$ (included) which is an average of 10 ensembles to save computational time.}\label{fig:wattstrogatz_complex}
\end{figure}
\clearpage
\subsubsection{Average degree = 5}
\begin{figure}[!htb]
    \includegraphics[width=1\textwidth, trim={0 1.5in 0 0in}]{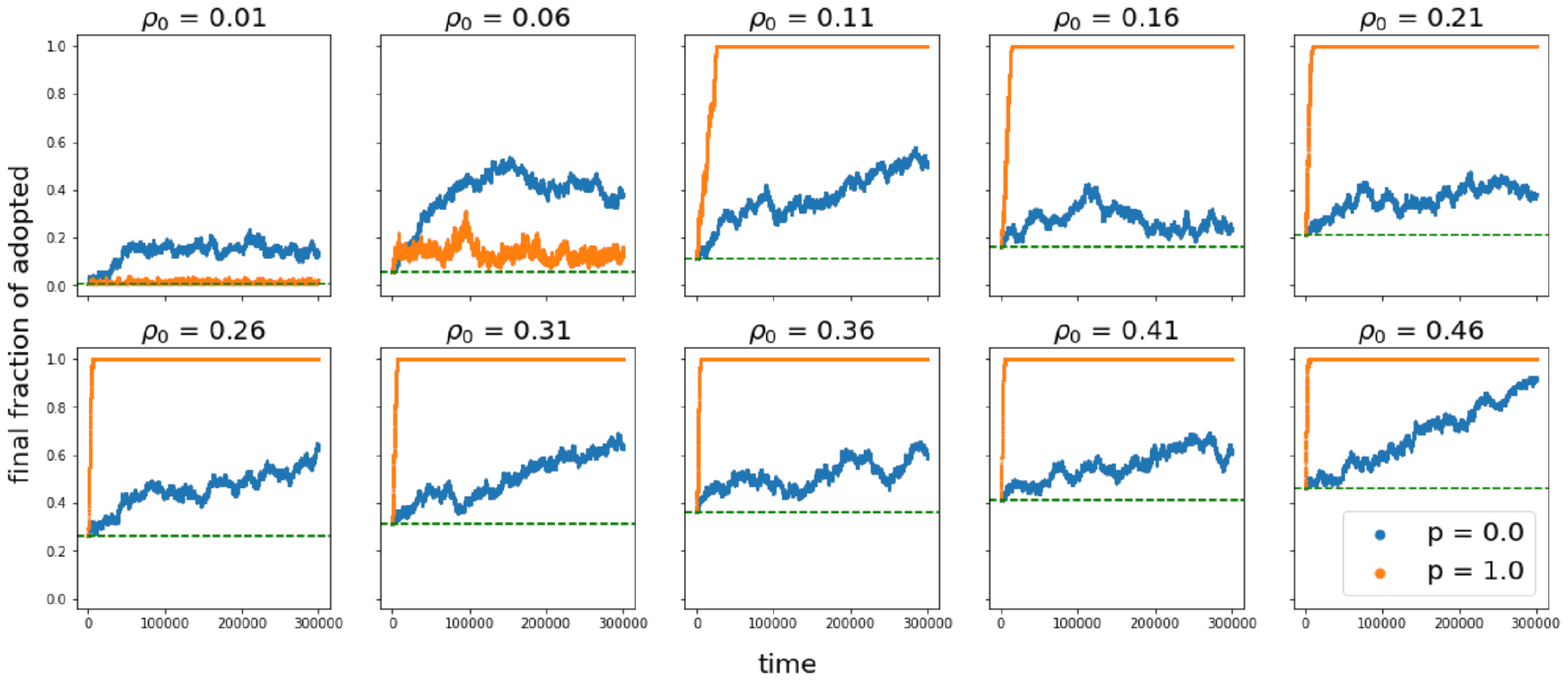}
    \caption{\textbf{The time evolution of the fraction of stable individuals of the type $\{+1,+1,+1\}$ in Scenario 2 of the Watts-Strogatz network setting.} We see a rich range of behaviors. In the case of $\rho_0=0.06$, we see that a clustered network structure (Watts Strogatz with rewiring probability $p=0$, blue curve) facilitates the contagion more than a random network structure (Watts Strogatz with rewiring probability $p=1$, orange curve) while in other cases it appears that a random network structure facilitates the contagion more than a highly clustered network structure. The horizontal green line indicates $\rho_0$. N=100, $\bar{k}=5, \sigma=0.2$. Curves are obtained from averaging $2$ ensembles for all subplots.}\label{fig:wattstrogatz_complex_k5}
\end{figure}
\clearpage

\section{Lack of an optimal regime of modularity for global diffusion of belief in a simple contagion setting}
Here is our analytical evidence that supports the lack of a regime of optimal network modularity in simple contagion.

Consider a susceptible-infected-recovered (SIR) model---a quintessential model of simple contagion. 
Let us explore the macroscopic dynamics of this model on the network with two communities using a mean-field approximation similar to the setup in~\citeauthor{nematzadeh2014optimal}[31]. 

Let us call the two communities as $A$ and $B$. 
Let $\Omega$ be the probability that a specific neighbor of an individual in $A$ is in $B$. 
Let $S_{A(B)}, I_{A(B)}, \eta_{A(B)}$ be the fraction of the susceptible, infected, and recovered individuals respectively, in community $A(B).$
The fraction of individuals in each of these compartments then evolve according to the following dynamical equations:
\begin{equation}~\label{eq:s}
    \frac{\partial S_{A(B)}}{\partial t} = -\beta[(1-\Omega) I_{A(B)} + \Omega I_{B(A)}]S_{A(B)},
\end{equation}
\begin{equation}~\label{eq:i}
    \frac{\partial I_{A(B)}}{\partial t} = \beta[(1-\Omega) I_{A(B)} + \Omega I_{B(A)}] - \gamma I_{A(B)},
\end{equation}
\begin{equation}~\label{eq:eta}
    \frac{\partial \eta_{A(B)}}{\partial t} = \gamma I_{A(B)},
\end{equation}
where $\beta$ is a the transmission rate between susceptible and infected individuals and $\gamma$ is the spontaneous recovery rate of infected individuals.   

Solving this system of equations based on the initial conditions that a fraction $\rho_0$ of individuals in $A$ are infected, i.e $S_A(0) = 1-\epsilon,$ $I_A(0) = \epsilon,$ $\eta_A(0) = 0$, and $S_B(0) = 1,$ $I_B(0) = 0,$ $\eta_B(0) = 0$, the fraction of recovered individuals (end state of the system since recovered individuals don't go back to being susceptible). 
At $t=\infty$, $\eta_A(\infty)$ and $\eta_B(\infty)$ can be derived as follows,

From Eq.~\ref{eq:eta},
\begin{equation}
    \int_0^{\infty}\frac{\partial\eta_A}{\partial t}dt = \eta_A(\infty) - \eta_B(0) = \eta_A(\infty) = \gamma\int_0^{\infty}I_A(t)dt~\label{eq:eta_sol}
\end{equation}

From Eqs.~\ref{eq:s} and~\ref{eq:eta_sol},
\begin{equation}~\label{eq:s_sol}
    -\frac{\gamma}{\beta}\int_0^{\infty}\frac{1}{S_A}\frac{\partial S_A}{\partial t}dt = (1 - \Omega)\gamma\int_0^\infty + \Omega\gamma\int_0^{\infty}I_B(t)dt = (1 - \Omega)\eta_A(\infty) + \Omega\eta_B(\infty).
\end{equation}

From the LHS of Eq.~\ref{eq:s_sol}
\begin{equation}
    -\frac{\gamma}{\beta}\int_0^{\infty}\frac{1}{S_A}\frac{\partial S_A}{\partial t}dt = -\frac{\gamma}{\beta}ln\frac{S_A(\infty)}{ S_A(0)} = -\frac{\gamma}{\beta}ln\frac{1-\eta_A(\infty)}{1-\rho_0} = (1 - \Omega)\eta_A(\infty) + \Omega\eta_B(\infty).
\end{equation}

Therefore, 
\begin{equation}
    \eta_A(\infty) = 1- (1-\rho_0)e^{-\frac{\beta}{\gamma}[(1-\Omega)\eta_A(\infty) + \Omega\eta_B(\infty)]}
\end{equation}

Similarly, based on the condition that $\eta_B = 0$,
\begin{equation}
    \eta_B(\infty) = 1 - e^{-\frac{\beta}{\gamma}[(1-\Omega)\eta_B(\infty) + \Omega\eta_A(\infty)]}
\end{equation}  

\begin{figure}[t]
    \centering
    \includegraphics[width=0.6\textwidth, trim={0 3in 0 3in}]{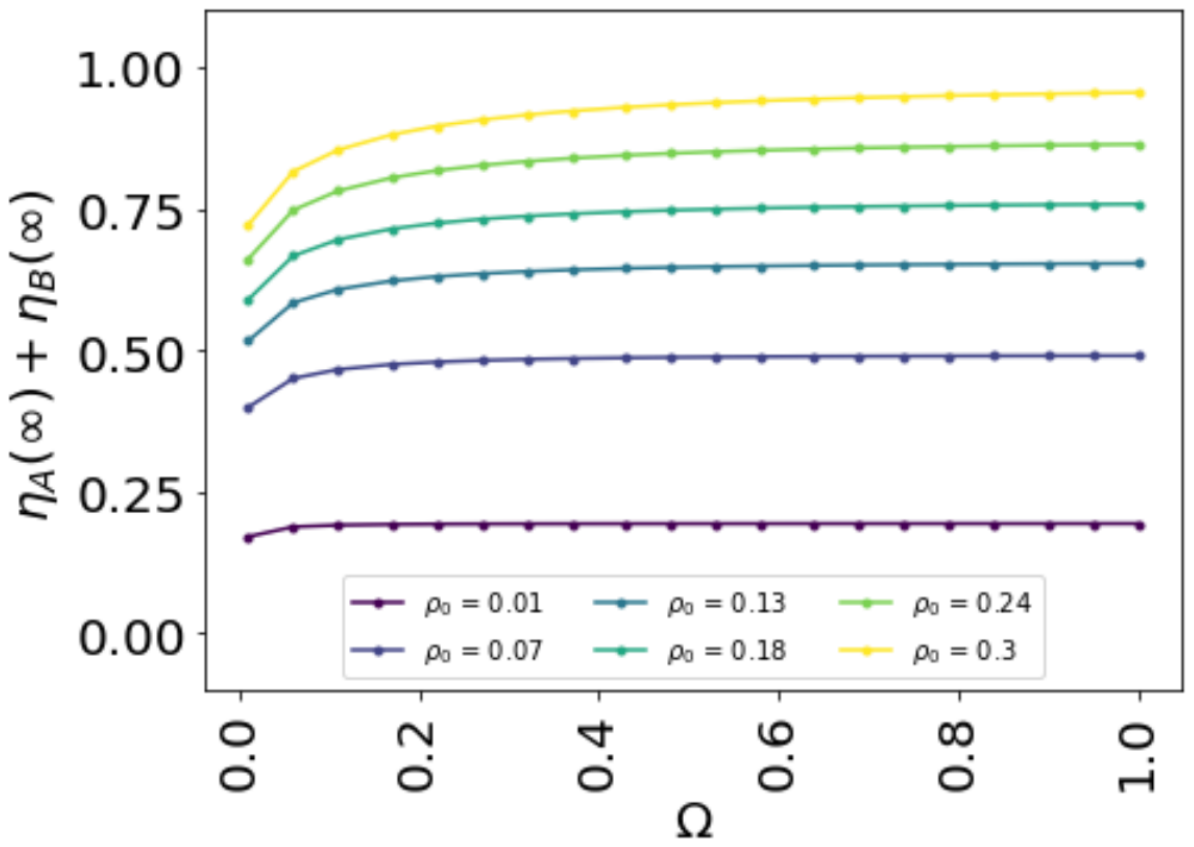}
    \caption{\textbf{Simple contagion setting does not show the optimal modularity behavior.} The total fraction of recovered individuals ($\eta_A(\infty) +\eta_B(\infty)$; range of [0,2])) in the mean field setup as probability ($\Omega$) of cross connection between two communities in the stochastic block model setup increases for different initial fraction of infected individuals ($\rho_0$). $\gamma = \beta = 1$.}\label{fig:sir_meanfield}
\end{figure}

Fig.~\ref{fig:sir_meanfield} shows that there is no optimal modularity. 
The results are qualitatively similar for different values of $\rho_0$, $\beta$ and $\gamma$.
\clearpage

\end{document}